\documentstyle[aps, prl, twocolumn, graphics, epsfig,
floats]{revtex}

\begin{document}

\title{Optimal Quantum Cloning via Stimulated Emission.
}

\author{Christoph Simon, Gregor Weihs, and Anton Zeilinger}

\address {Institut f\"ur Experimentalphysik, Universit\"at Wien,
Boltzmanngasse 5,\\ A-1090 Wien,  Austria\\ }

\date{August 12, 1999}
\maketitle

\begin{abstract}
We show that optimal universal quantum cloning can be realized via
stimulated emission. Universality of the cloning procedure is
achieved by choosing systems that have appropriate symmetries. We
first discuss a scheme based on stimulated emission in certain
three-level systems, e.g. atoms in a cavity. Then we present a way
of realizing optimal universal cloning based on stimulated
parametric down-conversion. This scheme also implements the
optimal universal NOT operation.
\end{abstract}


%\vskip 1pc

\narrowtext

It is not possible to construct a device that produces an exact
copy of an arbitrary quantum system \cite{wzurek}. This
impossibility has deep roots. It can be seen as a consequence of
the linearity of quantum mechanics. It also prevents the use of
EPR correlations for superluminal signaling \cite{gisin,herbert}.
Non-perfect copying, or cloning, though, is possible. Since the
seminal paper of Bu\v{z}ek and Hillery \cite{buzek96}, quantum
cloning has been extensively studied theoretically. Upper bounds
for the possible fidelity of quantum cloners have been derived
\cite{bruss}, and optimal universal quantum cloning
transformations have been discovered \cite{gisinmassar}.

All devices proposed so far consist of several quantum gates. This
means that it will probably take some time until their practical
realization. On the other hand, cloning was originally discussed
in the context of stimulated emission \cite{herbert,wzurek}. It
was realized that {\it perfect} copying is prevented by the
unavoidable presence of spontaneous emission \cite{milonnimandel}.
The question arises whether {\it optimal} cloning (for which the
fidelity of the clones saturates the above-mentioned bounds) can
be realized with stimulated emission. In this letter, suggesting
realistic scenarios, we show that the answer is yes.

The cloning procedure will clearly be universal, i.e. equally good
for all possible input states, if the cloning system is symmetric
under general unitary transformations of the system that is to be
cloned. To be more specific, consider cloning of a general qubit
represented by the polarization state of a photon. This requires a
population inverted medium whose initial state and whose
interaction Hamiltonian with the electromagnetic field are both
invariant under general polarization transformations so that it
can emit photons of any polarization with the same probability. If
a photon enters such a medium, it stimulates the emission of
photons of the same polarization.

Of course, there is also spontaneous emission of photons of the
wrong polarization. The presence of this spontaneous emission is
unavoidable because for a given transition the total emission
amplitude (i.e. stimulated and spontaneous) is $\sqrt{n+1}$ times
the amplitude for spontaneous emission, where $n$ is the number of
stimulating photons present. As we want stimulated emission to be
possible for all photon polarizations in order to achieve
universality, this means that spontaneous emission will also occur
for all polarizations.

The photons in the final state can be considered as clones of the
original incoming photon. The fidelity of the clones is limited by
the presence of the spontaneously emitted photons. The fidelity is
defined as $\langle\psi|\rho^{out}|\psi\rangle$, where
$|\psi\rangle$ is the state of the original qubit and $\rho^{out}$
is the reduced density matrix of one of the clones. This is
equivalent to the relative frequency of photons of the right
polarization in the final state.  Starting from one qubit, an
optimal universal symmetrical cloner \cite{bruss} produces $M$
identical clones with a fidelity $F_{opt}(M)=
\frac{2}{3}+\frac{1}{3M}$. Note that $M=2$ which gives
$F_{opt}=5/6$ means that there is just one additional qubit
besides the original.

\begin{figure}
   \begin{center}\mbox{\input epsf \epsfxsize0.3 \columnwidth
\epsfbox{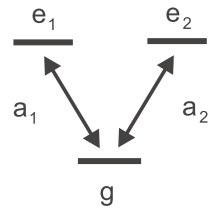}}\end{center}
        \caption{One possible level structure of systems used for universal cloning, optimal for
        short interaction times.}

\end{figure}

The first possible practical realization of quantum cloning we
discuss is based on stimulated emission in an ensemble of
three-level systems. These systems
 have a
ground level $g$ and two degenerate upper levels $e_1$ and $e_2$,
connected by two orthogonal modes of the electromagnetic field,
$a_1$ and $a_2$ (see fig.1). The field modes define the Hilbert
space of our qubits, i.e. we want to clone general superposition
states $(\alpha a^\dagger_1 + \beta a^\dagger_2) |0\rangle$.  Note
that we are talking about photons and polarization in order to be
specific, but one is free to think of other systems and other
degrees of freedom, as long as they are described by the same
formalism. In the interaction picture, the Hamiltonian has the
following form:
\begin{equation}
H=\sum \limits_{K=1}^N \gamma ( \sigma^K_{+1} a_1 + \sigma^K_{+2}
a_2) + h.c. \label{hatoms}
\end{equation}
where $\sigma_{+1(2)}=|e_{1(2)}\rangle \langle g|$, its complex
conjugate is denoted by $\sigma_{-1(2)}$, and the index $K$ refers
to the $K$th atom.

\begin{figure}

        \includegraphics[width=  \columnwidth] {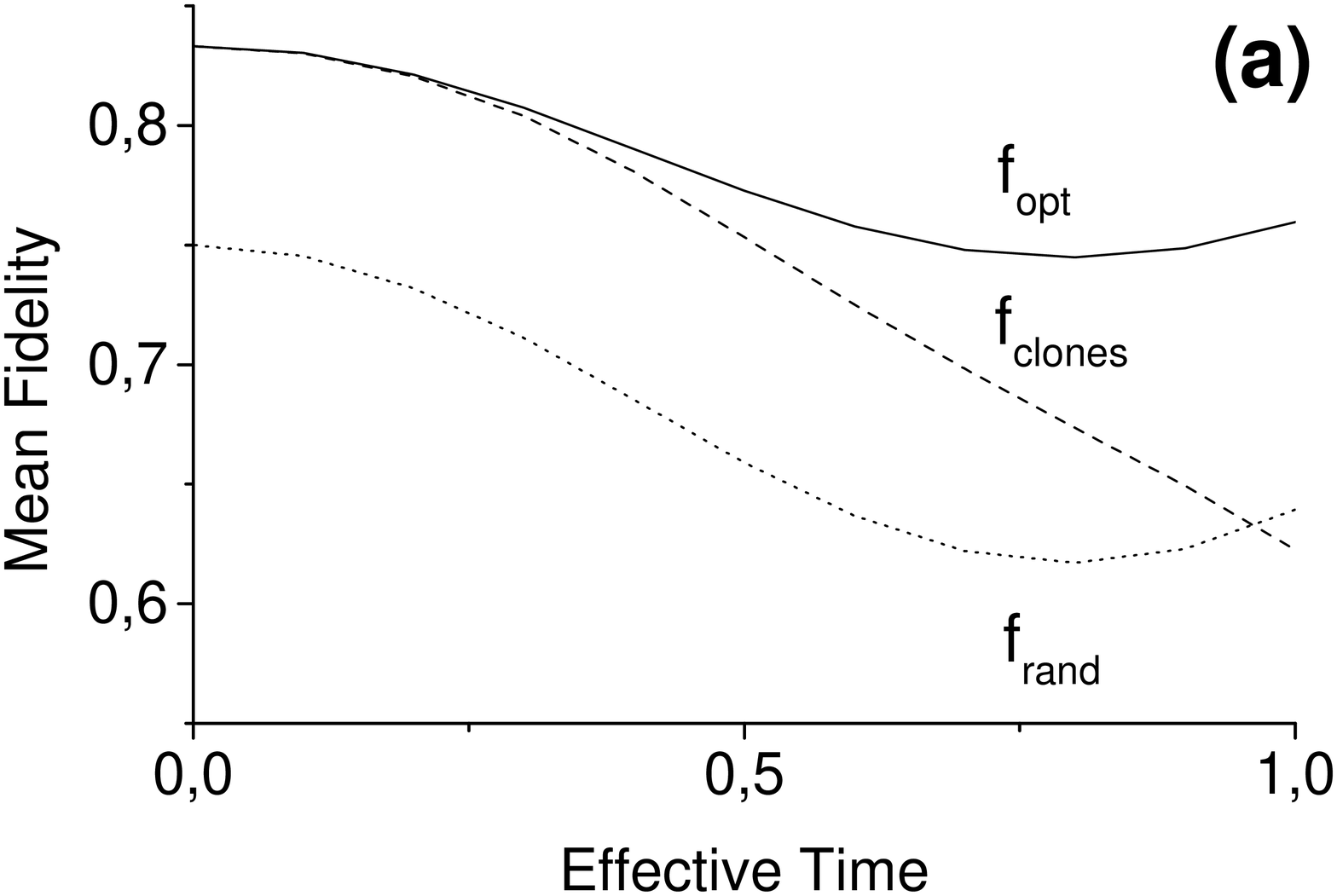}
        \includegraphics[width=  \columnwidth] {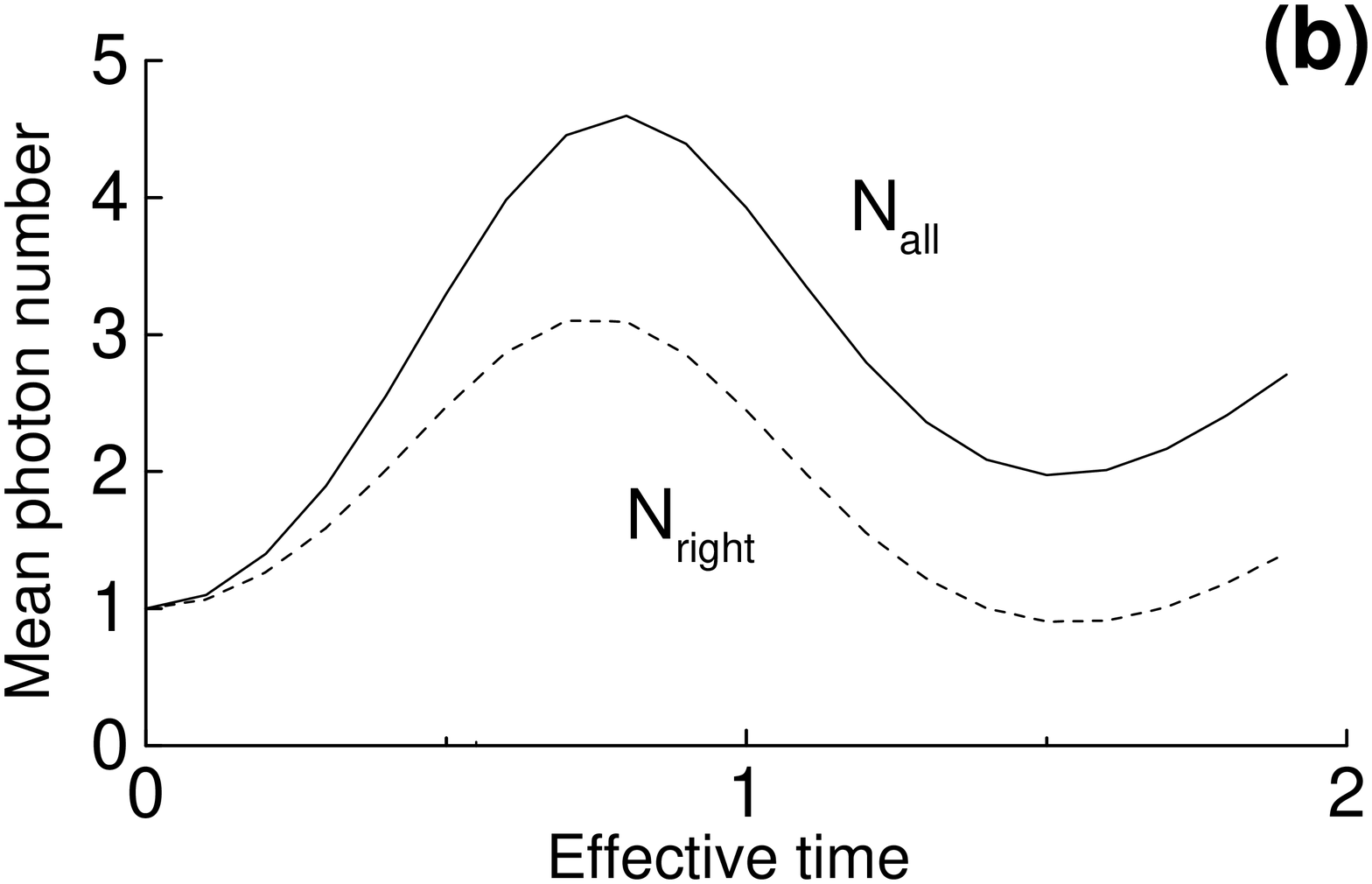}
        \caption{(a) Dependence on time, measured in units of $\gamma t$, of $f_{opt}$, $f_{clones}$,
         and $f_{rand}$, which are the optimum possible fidelity, the
         fidelity achieved by our three-level cloning procedure, and the fidelity achieved by random photon
        production respectively, as defined in eq. (3-5) for the case of $N=6$ atoms.
        It is evident that optimal cloning is achieved in the short-time
        limit. (b) Time dependence of the mean number
        of all photons $N_{all}$ and of the mean number of ``right''
        photons (i.e. of the same polarization as the incoming
        photon)
        $N_{right}$ for the case $N=6$.
        }
        \label{f:powershade}
\end{figure}

The Hamiltonian (\ref{hatoms}) is invariant under simultaneous
unitary transformations of the vectors $(a^\dagger_1,
a^\dagger_2)$ and $(|e_1\rangle, |e_2\rangle)$. Furthermore, we
require each atom to be initially in the mixed state
\begin{equation}
\rho_{i}=\frac{1}{2} (|e_1\rangle \langle e_1| +|e_2\rangle
\langle e_2| )
\end{equation}
which is invariant under the same unitary transformations. The
invariance of both Hamiltonian and initial state together ensure
the universality of the cloning procedure. Therefore it is
sufficient to analyze the performance of the cloner for one
arbitrary incoming one-photon state, e.g.
$|\psi_i\rangle=a^\dagger_1 |0\rangle$.

 We have performed numerical computations for
systems of a few (up to $N=6$) atoms.  From (\ref{hatoms}), the
time development operator $U=e^{-iHt}$ for the whole atoms-photons
system was calculated. Use was made of the fact that $N_1$ and
$N_2$, which denote the sum of the number of photons plus the
number of excited atoms for mode 1 and 2 respectively, are
independently conserved quantities. Therefore the whole Hilbert
space is decomposable into invariant subspaces, i.e. $H$ and $U$
are block-diagonal.

The final state of the procedure has components with various
numbers of photons, where the maximum total number is $N+1$ (if
all atoms have emitted their photons). The probability to find $k$
``right'' and $l$ ``wrong'' photons in the final state, denoted by
$p(k,l)$, was calculated for all possible values of $k$ and $l$
and for different values of $\gamma t$, and from it the overall
average ``fidelity''

\begin{equation}
f_{clones}\!=\! \sum \limits_{k+l\geq2} p'(k,l) \left(
\frac{k}{k+l} \right)
\end{equation}
was determined. This is the average of the relative frequency of
photons with the correct polarization in the final state. The
average is performed only over those cases where there are at
least two photons in the final state, i.e. where at least one
clone has been produced. $p'(k,l)=p(k,l)/(1-p(1,0)-p(0,1))$ is
used in order to have proper normalization. Note that $p(0,0)$ is
always zero.

That average fidelity for our cloning procedure was compared to
the average fidelity that would be achieved by
 an ensemble of optimal
cloners producing the same distribution of numbers of photons,
i.e. to
\begin{equation}
f_{opt}\!=\! \sum \limits_{n=2}^{N+1} p'(n)\left( \frac{2n+1}{3n}
\right),
\end{equation}
where $p'(n)\!=\!\sum \limits_{k+l=n} p'(k,l)$. We also made a
comparison to the case, where, in addition to the incoming photon,
photons are just created randomly, i.e. to the fidelity
\begin{equation}
f_{rand}\!=\!\sum \limits_{n=2}^{N+1} p'(n) \left( \frac{n+1}{2n}
\right) .
\end{equation}

Figure 2 shows that the fidelity of our cloning procedure
approaches the optimum fidelity for early times. For short times
the probability for every individual atom to have already emitted
its photon is low. The time behaviour of the mean number of
photons and also of the mean number of photons of the correct
polarization produced is shown in fig. 2(b). Therefore, in order
to produce a reasonable average number of clones in this regime, a
large number of atoms is necessary.

The practical realization of this scheme probably requires a
cavity in order to achieve the interaction of a single spatial
mode of the radiation field with several (or even many) atoms.
Trapping several atoms in a cavity could be possible. The atoms
could also fly through the cavity \cite{flythru}.

The second scheme for quantum cloning that we want to present is
based on stimulated parametric down-conversion (PDC). We will show
that optimal cloning can be realized. In PDC a strong light beam
is sent through a crystal. There is a certain (very low)
probability for a photon from the beam to decay into two photons
such that energy and crystal momentum are conserved. In type-II
PDC the two photons that are created have different polarization.
They are denoted as signal and idler.

Figure 3 shows the setup that we have in mind. We consider pulsed
type-II  frequency-degenerate PDC. It is possible to choose two
conjugate directions for the signal and idler beams such that
photon pairs that are created along these two directions are
entangled in polarization. This source of
polarization-entanglement \cite{source} has been used in many
experiments \cite{telep}. We consider the quasi-collinear case
(i.e. the two directions almost coincide), so that the transverse
motion of the photons in the crystal is not important.

\begin{figure}
   \begin{center}\mbox{\input epsf \epsfxsize0.9 \columnwidth
\epsfbox{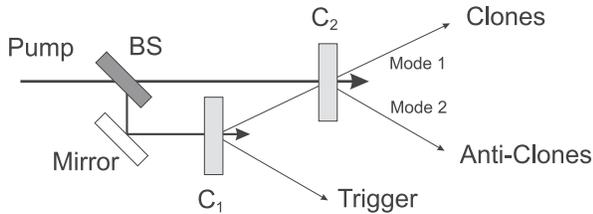}}\end{center}
        \caption{Setup for optimal cloning by parametric down-conversion
        { \protect \cite{Mandel,DeM,Migdall}}.
The pump-pulse is split at the beam splitter BS. One part of the
pump pulse hits the first crystal C$_1$, where photon pairs are
created with a certain rate.
 One photon from  each pair can be used as a trigger.  The other photon is
the system to be cloned. This photon is directed towards the
second crystal C$_2$, where it stimulates emission of photons of
the same polarization along the same direction. The path lengths
have to be adjusted in such a way that the DC-photon and the
second part of the pump pulse reach C$_2$ simultaneously. The
photons in mode 1 are optimal clones of the incoming photon, and
the photons in mode 2 are the output of an optimal universal
NOT-gate. It is interesting to note that in this scheme one is
actually cloning a photon that is part of an entangled pair. }
\label{pdc}
\end{figure}

For stimulated emission to work optimally, there has to be maximum
overlap of the amplitudes of the incoming photon and of all the
photons that are produced in the second crystal. This can be
achieved by using a pulsed scheme together with filtering of the
photons before detection \cite{ghzvorschlag}. The pump pulse can
be seen as an active volume that moves through the crystal. If the
photons are filtered so much that the smallest possible size of
the wavepackets detected is substantially bigger than the pump
pulse, then there is maximum overlap between different pairs
created in the same pulse.  Of course, filtering limits the
achievable count rates. Moreover the group velocities of pump
pulse, signal ($V$) and idler ($H$) photons are not all identical.
This leads to separations (of the order of a few hundred fs per
millimeter in BBO), which have to be kept small compared to the
size of the DC-photon wave packets. There is a trade-off between
filtering and crystal length, i.e. one can choose narrower filters
in order to be able to use a longer crystal (which leads to longer
interaction times).

If the above-mentioned conditions are fulfilled, then a single
spatial mode (i.e. one mode for the signal and one for the idler
photons) approximation can be used. The PDC process can then be
described in the limit of a large classical pump pulse, in the
interaction picture, by the Hamiltonian
\begin{equation}
H=\gamma(a^\dagger_{V1} a^\dagger_{H2} - a^\dagger_{H1}
a^\dagger_{V2}) + h.c.,
\end{equation}
where $a^\dagger_{V1}$ is the creation operator for a photon with
polarization V propagating along direction 1 etc. The coupling
constant and the intensity of the classical pump pulse are
contained in $\gamma$.

The Hamiltonian $H$ is invariant under general common $SU(2)$
transformations of the polarization vectors $(a^\dagger_V,
a^\dagger_H)$ for modes $1$ and $2$, while a phase transformation
will only change the phase of $\gamma$. This makes our cloner
universal, i.e. its performance is polarization independent.
 Therefore it is
sufficient to analyze the ``cloning'' process in one basis.

The time development operator $e^{-iHt}$ clearly factorizes into a
$V1-H2$ and an $H1-V2$ part. Consider cloning starting from $N$
identical photons in the initial state
$|\psi_i\rangle=\frac{(a^\dagger_{V1})^N}{\sqrt{N!}} |0\rangle$
Making use of the disentangling theorem \cite{Walls} one finds
that (cf. \cite{DeM})
\begin{eqnarray}
|\psi_f\rangle=e^{-iHt}|\psi_i\rangle=K& & \sum
\limits_{k=0}^{\infty}(-i \Gamma)^k\sqrt{{k+N \choose N}
}|k+N\rangle_{V1}|k\rangle_{H2}\nonumber\\ \times &
&\sum\limits_{l=0}^{\infty} (i
\Gamma)^l|l\rangle_{H1}|l\rangle_{V2} \label{finalstate}
\end{eqnarray}
where $\Gamma=\tanh \gamma t$ and $K$ is a normalizing factor.

The component of this state which has a fixed number $M$ of
photons in mode 1, is proportional to
\begin{equation}
\sum \limits_{l=0}^{M-N} (-1)^l \sqrt{{M-l \choose N}}
|M-l\rangle_{V1} |l\rangle_{H1} |l\rangle_{V2} |M-N-l\rangle_{H2}.
\label{subspacestate}
\end{equation}
This is identical to the state produced by the unitary
transformation written down in \cite{NOT} which can be seen as a
special version of the Gisin-Massar cloners \cite{gisinmassar}
that implements optimal universal cloning and the optimal
universal NOT-gate at the same time. The $M$ photons in mode 1 are
the clones, while the $M-N$ photons in mode 2 are the output of
the universal NOT-gate, the ``anti-clones''.

In order to see that state (\ref{subspacestate}) is indeed the
output of an optimal cloner, let us calculate the relative
frequency of photons of the ``right'' polarization in mode 1. It
is given by
\begin{equation}
f_{clones}^N(M)=\frac{\sum \limits_{l=0}^{M-N} {M-l \choose
N}(M-l)}{M \sum \limits_{l=0}^{M-N} {M-l \choose N}}.
\end{equation}

Using $\sum \limits_{k=N}^M {k \choose N} = {M+1 \choose N+1}$ it
follows that

\begin{equation}
f_{clones}^N(M)=\frac{NM+N+M}{M(N+2)},
\end{equation}

which is exactly the optimum fidelity for an $N$ to $M$ quantum
cloner \cite{bruss}. A similar calculation demonstrates that the
universal NOT is realized in mode 2.

This means that the setup of fig. \ref{pdc} works as an ensemble
of optimal universal cloning (and universal NOT) machines,
producing different numbers of clones and anti-clones with certain
probabilities. Note that each of the modes can be used as a
trigger for the other one and therefore cloning or anti-cloning
with a fixed number of output-systems can be realized by
post-selection.

We have shown a method of realizing optimal quantum cloning
machines. We emphasize that this scheme should be experimentally
feasible with current technology. In our group, pair production
probabilities of the order of $4 \cdot 10^{-3}$ have been achieved
with a 76 MHz pulsed laser system (UV-power about 0,3 W) and a 1
mm BBO crystal, for 5 nm filter bandwidth. Past experiments show
that good overlap of photons originating from different pairs is
achieved under these conditions. With detection efficiencies
around 10 percent, this leads to a rate of two-pair detections of
the order of one per a few seconds.

A new 300 kHz laser system is currently being set up in our lab.
An improvement of the order of $\frac{76}{0,3}$  in the average
rate of pairs per pulse is to be expected, for identical pump
power. This will also make several-pair events far more likely.
This means that production of a few clones with a reasonable rate
should be possible.

Here we have presented possible ways of realizing quantum cloning
via stimulated emission. We have first discussed a procedure based
on three-level systems that could allow the production of large
numbers of clones, and could be easier to realize than comparable
schemes using quantum gates. Then we have shown a scheme for
realizing optimal universal cloning based on parametric
down-conversion. This scheme should be realizable with current
technology.

We would like to thank S. Stenholm for a provocative question that
triggered or re-animated our interest in the relationship between
quantum cloning and stimulated emission, and  V. Bu\v{z}ek for
helpful comments, in particular for bringing the connection
between cloning and the universal NOT to our attention. We also
thank \v{C}. Brukner, J.I. Cirac, T. Jennewein, J.W. Pan, and  H.
Weinfurter for helpful comments. This work has been supported by
the Austrian Science Foundation (FWF, Projects No. S6502 and
F1506).

\end{document}